\documentclass[12pt]{article}

\usepackage{epic}
\usepackage{eepic}
\usepackage{graphics}
\usepackage{amsfonts}
  \usepackage{amsthm}
 \usepackage{amsmath}
 \usepackage{amscd}
 \usepackage{amssymb}
 \usepackage{latexsym}
\usepackage{graphicx}

 \numberwithin{equation}{section}
 \newtheorem{thm}{Theorem}
 
 \newtheorem{prop}{Proposition}
 \theoremstyle{definition}

\title{A note on the integrability of Hamiltonian 1 : 2 : 2 resonance}
\author{O. Christov\\
Faculty of Mathematics and Informatics, Sofia University, \\
5 J. Bouchier blvd., 1164 Sofia, Bulgaria}
\date{}

\begin{document}

\maketitle

\begin{abstract}
\noindent
We study the integrability of the Hamiltonian normal form of 1 : 2 : 2 resonance.
It is known that this normal form truncated to order three is integrable.
The truncated  to order four normal form contains too many parameters.
For a generic choice of parameters in the normal form up to order four we prove
a non-integrability result using Morales-Ramis theory.
We also  isolate a non-trivial case of integrability.
\end{abstract}

{\bf Keywords:} Hamiltonian 1 : 2 : 2 resonance, Liouville integrability Differential Galois groups, Morales-Ramis theory

\section{Introduction}

For an analytic Hamiltonian $H (q, p)$ with an elliptic equilibrium at the origin,
we have the following expansion
\begin{equation}
\label{expan}
H = H_2 + H_3 + H_4 + \ldots,
\end{equation}
where
$
H_2 = \sum \omega_j (q_j^2 + p_j^2), \quad \omega_j > 0
$
and $H_j$ are homogeneous of degree $j$.

It is said that the frequency vector $\omega = (\omega_1, \ldots, \omega_n)$ satisfies a
resonant relation if there exists a vector $k = (k_1 \ldots, k_n), k_j \in \mathbb{Z}$, such that
$(\omega, k) = \sum k_j \omega_j = 0$, $| k | = \sum | k_j |$ being the order of the resonance.

There exists a procedure called normalization, which simplifies the Hamiltonian function in a neighborhood of the equilibrium and
and is achieved by means of canonical near-identity transformations \cite{AKN,SaVerM,V1}.
When resonances appear this simplified Hamiltonian is called Birkhoff-Gustavson normal form. To study the behavior of
a given Hamiltonian system near the equilibrium one usually considers the normal form truncated to some order
\begin{equation}
\label{truncation}
\bar{H} = H_2 + \bar{H}_3 + \ldots + \bar{H}_m .
\end{equation}
Note that by construction $\{\bar{H}_j, H_2\} = 0$ ( $\{ , \} $  being the  Poisson bracket). This means that the truncated
resonant normal form has at least two integrals - $\bar{H}$ and $H_2$.

The first integrals for the resonant normal form $\bar{H}$ are approximate integrals for the original system,
see Verhulst \cite{V1} for the precise statements. If the truncated normal form happens to be
integrable then the original Hamiltonian system is called near-integrable.

In this note we study the integrability of the semi-simple Hamiltonian  1 : 2 : 2 resonance.
The classical water molecule model and concrete models of coupled rigid bodies
serve as examples  which are described by the Hamiltonian systems in 1 : 2 : 2 resonance,
see Haller \cite{Haller}.

A recent review of some known results on integrability of
the Hamiltonian normal forms can be found in \cite{V2}.

When studying normal forms it is customary to introduce the complex coordinates
$$
z_j = q_j + i p_j, \quad \zeta_j = q_j - i p_j .
$$
The generating functions of the algebra of the elements which are Poisson-commuting with
$$
H_2 = (q_1^2 + p_1^2) + 2 (q_2^2 + p_2^2) + 2 (q_3^2 + p_3^2)
$$
are written, for example  in \cite{Hansmann,SaVerM}:
\begin{eqnarray}
\label{generators}
 & z_j \zeta_j, \, \, j = 1, 2, 3, \quad z_2 \zeta_3, \quad \zeta_2 z_3 ,  \nonumber \\
 & z_1^2 \zeta_2, \quad  \zeta_1^2 z_2, \quad z_1^2 \zeta_3, \quad  \zeta_1^2 z_3 .
\end{eqnarray}
Then the Hamiltonian normal form up to order 4 is
\begin{equation}
\label{nform4}
\bar{H} = H_2 + \bar{H}_3 + \bar{H}_4,
\end{equation}
where
\begin{align}
\label{1.1}
H_2       & = z_1 \zeta_1 + 2 z_2 \zeta_2 +  2 z_3 \zeta_3 , \nonumber \\
\bar{H}_3 & = a z_1 ^2 \zeta_2 + \bar{a} \zeta_1 ^2 z_2 + b z_1 ^2 \zeta_3 + \bar{b} \zeta_1 ^2 z_3 ,  \\
\bar{H}_4 & = c_1 (z_1 \zeta_1)^2 + c_2 z_1 \zeta_1 z_2 \zeta_2 + c_3  z_1 \zeta_1 z_3 \zeta_3  + c_4 z_2 \zeta_2 z_3 \zeta_3
     +  c_5 (z_2 \zeta_2)^2  +  c_6 (z_3 \zeta_3)^2  + \nonumber \\
 & d (z_2 \zeta_3)^2  + \bar{d} (\zeta_2 z_3)^2
  + (z_1 \zeta_1)(e z_2 \zeta_3 + \bar{e} \zeta_2 z_3) + (z_2 \zeta_2)(f z_2 \zeta_3 + \bar{f} \zeta_2 z_3)
 + (z_3 \zeta_3)(g z_2 \zeta_3 + \bar{g} \zeta_2 z_3) .    \nonumber
\end{align}

For the normal form of the 1 : 2 : 2 resonance, normalized to degree three
\begin{equation}
\label{degr3}
\bar{H} = H_2 + \bar{H}_3
\end{equation}
symplectic coordinate changes allow to make the coefficients in $\bar{H}_3$ real and additionally to achieve
$b = 0$ \cite{AaV1,Kummer}. This makes the cubic normal form integrable
(see also \cite{Aa1} where results  on the integrability for other first order resonances are given).
In particular, detailed geometric analysis
based on  1 : 2 resonance is given in \cite{Kummer,HW,Haller}.
To see what happens in the normal form normalized to order four, van der Aa and Verhulst \cite{AaV1} consider
 a specific term from $\bar{H}_4$, which destroys the intrinsic symmetry of (\ref{degr3}) and as a result
causes the loss of the extra integral.

More geometric approach is taken by Haller and Wiggins \cite{HW,Haller}. For a class of resonant
Hamiltonian normal forms (1 : 2 : 2 among them) normalized to degree four, they prove the following results.
First, most invariant 3-tori of the cubic normal form (\ref{degr3}) survive on all but finite number of energy
surfaces. Second, there exist wiskered 2-tori which intersect in a non-trivial way giving rise to multi-pulse
homoclinic and heteroclinic connections. The existence of these wiskered 2-tori is the "geometric" source of
non-integrability.

To study the integrability of the normal form (\ref{1.1}) we adopt more algebraic approach here.
We assume that at least one of $(a, b)$ is different from zero, say $a \neq 0, b = 0$ (see the above explanations).
On the contrary, if $a = b = 0$, i.e., there is no cubic part, there exists an additional integral
$I_1 := z_1 \zeta_1 = p_1 ^2 + q_1 ^2$ and the normal form truncated to order 4 is integrable.
In particular, the normal form
\begin{align}
\label{1.2}
\bar{H} & = \frac{1}{2} (p_1^2 + q_1^2) + \frac{2}{2}(q_2^2 + p_2^2 + q_3^2 + p_3^2) \\
 & + a_1 (p_1^2 + q_1^2)^2 + a_2 (p_2^2 + q_2^2 + p_3^2 + q_3^2)^2 + a_3 (p_1^2 + q_1^2)(p_2^2 + q_2^2 + p_3^2 + q_3^2) +
     +  a_4 (p_3 q_2 -q_3 p_2)^2  \nonumber
\end{align}
is integrable with the following quadratic first integrals
\begin{equation}
\label{1.3}
I_1 := p_1^2 + q_1^2, \quad F_1 := p_2^2 + q_2^2 + p_3^2 + q_3^2, \quad F_2 := p_3 q_2 - q_3 p_2 .
\end{equation}
For this case, the action-angle variables are introduced in a similar way as in \cite{Rink} and
the KAM theory conditions can be verified upon certain restrictions on the coefficients $a_j$
of the normal form (\ref{1.2}).

Further, we perform a time dependent canonical transformation as in \cite{Ford,OC2} to eliminate the quadratic part of
(\ref{1.1}). Still there are too many parameters in (\ref{1.1}), so we take a generic choice which in
 cartesian coordinates reads
\begin{align}
\label{1.10}
\bar{H} & =  a [q_2 (q_1 ^2 - p_1 ^2) + 2 q_1 p_1 p_2] \\
 & + \frac{\alpha}{2} [ (q_2 q_3 + p_2 p_3)^2 - (p_2 q_3 - q_2 p_3)^2] + \frac{\beta}{2}(q_2 q_3 + p_2 p_3) (q_1 ^2 + p_1 ^2) . \nonumber
\end{align}

The main result of this note is the following
\begin{thm}
\label{th1}
For the  system governed by the Hamiltonian (\ref{1.10}) we have the following assertions:

(i) if $\alpha \neq 0$ it is non-integrable;

(ii) if $\alpha = 0$ it is Liouville integrable with the additional first integral
\begin{equation}
\label{fint}
F = 4 a^2 (q_3^2 + p_3^2) - a \beta \left[q_3 (q_1^2 - p_1^2) + 2 q_1 p_1 p_3 \right]+
\frac{\beta^2}{16}(q_1^2 + p_1^2)^2 + \beta^2(q_2 q_3 + p_2 p_3)^2 .
\end{equation}
\end{thm}

In order to prove (i) we make use of Morales-Ramis theory, see \cite{M,MRS1,MR2} for the fundamental theorems and examples of their applications.
This theory gives necessary conditions for Liouville integrability of a Hamiltonian system in terms of abelianity  of the differential Galois
groups of variational equations along certain particular solution. It generalizes the Ziglin's work \cite{Ziglin} who
uses the monodromy group of the variational equations instead.

All the notions, definitions, statements and proofs about the differential Galois theory can be found in \cite{Mi,Magid,vPS,Singer}.

The proof of Theorem \ref{th1} is given in the next section.
We finish with some remarks on application of this approach in investigating the integrability of other Hamiltonian resonances.

\section{Proof of Theorem \ref{th1}}

In this section we prove Theorem \ref{th1}. The proof of (i) is carried out in two steps. First, we find a particular solution and
write the variational equation along it. It appears that its Galois group is $\mathrm{SL} (2, \mathbb{C})$
except for the cases $\beta/(\sqrt{2} \alpha) \in \mathbb{Z}$. Then we take the simplest case $\beta = 0$ and choose another particular solution.
The Galois group of the corresponding variational equations turns out to be solvable, but non-commutative.

{\bf Proof:}
The Hamilton's equations corresponding to (\ref{1.10}) are
\begin{eqnarray}
\label{sys}
\dot{q}_1 & = & 2 a(q_1 p_2 - q_2 p_1) + \beta p_1 (q_2 q_3 + p_2 p_3), \nonumber \\
\dot{p}_1 & = & -2a(q_1 q_2 + p_1 p_2) - \beta q_1 (q_2 q_3 + p_2 p_3), \nonumber \\
\dot{q}_2 & = & 2 a q_1 p_1 + \alpha [ p_3 (q_2 q_3 + p_2 p_3) - q_3 (p_2 q_3 - q_2 p_3) ] + \frac{\beta}{2} p_3 (q_1^2 + p_1^2),  \\
\dot{p}_2 & = & -a (q_1^2 - p_1^2) - \alpha [q_3 (q_2 q_3 + p_2 p_3) + p_3 (p_2 q_3 - q_2 p_3) ] - \frac{\beta}{2} q_3 (q_1^2 + p_1^2),\nonumber \\
\dot{q}_3 & = & \alpha [ p_2 (q_2 q_3 + p_2 p_3) + q_2 (p_2 q_3 - q_2 p_3) ] + \frac{\beta}{2} p_2 (q_1^2 + p_1^2), \nonumber \\
\dot{p}_3 & = & - \alpha [q_2 (q_2 q_3 + p_2 p_3) - p_2 (p_2 q_3 - q_2 p_3) ] - \frac{\beta}{2} q_2 (q_1^2 + p_1^2) . \nonumber
\end{eqnarray}

In order to apply the Morales-Ramis theory we need a non-equilibrium solution.
\begin{prop}
\label{PS1}
Suppose $\alpha \neq 0$. Then the system (\ref{sys}) has a particular solution of the form
\begin{equation}
\label{4.1}
q_2  = \frac{1}{\sqrt{2 \alpha t}}, \quad q_3  = p_3  = \frac{i}{2 \sqrt{\alpha t}}, \quad q_1  = p_1  = p_2  = 0.
\end{equation}

\end{prop}
The proof is straightforward.

$\hfill \square$

Denote $dq_j = \xi_j, dp_j = \eta_j, j= 1,2,3$. Then the normal variational equations (NVE) along the solution (\ref{4.1})
is written in $\xi_1, \eta_1$ variables
\begin{equation}
\label{4.4}
\dot{\xi}_1   =  \left[ \frac{i \beta}{2 \sqrt{2} \alpha t} - \frac{\sqrt{2}a}{\sqrt{\alpha t}}\right] \eta_1  ,  \quad
\dot{\eta}_1  =  -\left[ \frac{i \beta}{2 \sqrt{2} \alpha t} + \frac{\sqrt{2}a}{\sqrt{\alpha t}}\right] \xi_1 .
\end{equation}
Further, we put $t = \tau^2$ (in fact, this transformation is a two-branched covering mapping which
 preserves the identity component of the Galois group of (\ref{4.4})).
 Denote $P:= \frac{i \beta}{\sqrt{2}\alpha}, Q:= \frac{2 \sqrt{2} a}{\sqrt{\alpha}} \neq 0$.
 Then the system (\ref{4.4}) becomes
\begin{equation}
\label{4.5}
\begin{pmatrix}
\xi_1 \\ \eta_1
\end{pmatrix}^{'} =
\left[
\begin{pmatrix}
0 & -Q \\
-Q & 0
\end{pmatrix}
+ \frac{1}{\tau}
\begin{pmatrix}
0 & P \\
-P & 0
\end{pmatrix}
\right]
\begin{pmatrix}
\xi_1 \\ \eta_1
\end{pmatrix}, \qquad (' = d/d \tau) .
\end{equation}
We scale the independent variable $\tau \to Q \tau$ and perform a linear change
$$
\begin{pmatrix}
\xi_1 \\ \eta_1
\end{pmatrix} = T
\begin{pmatrix}
x \\ y
\end{pmatrix}, \quad
T = \frac{1}{\sqrt{2}}
\begin{pmatrix}
1 & -1 \\
1 & 1
\end{pmatrix},
$$
 which transforms the leading matrix in (\ref{4.5}) into diagonal form. Thus we get
\begin{equation}
\label{4.6}
\begin{pmatrix}
x \\[1ex] y
\end{pmatrix}^{'} =
\left[
\begin{pmatrix}
-1 & 0 \\[1ex]
0 & 1
\end{pmatrix}
+ \frac{1}{\tau}
\begin{pmatrix}
0 & P \\[1ex]
-P & 0
\end{pmatrix}
\right]
\begin{pmatrix}
x \\[1ex] y
\end{pmatrix}.
\end{equation}
For the system (\ref{4.6}) $\tau = 0$ is a regular singular point and
$\tau = \infty$ is an irregular singular point.

Now we study the local Galois group $G_{\infty}$.
By a Theorem of Ramis \cite{MarRam,Mi} this group is topologically generated by the formal monodromy,
the exponential torus and the Stokes matrices. One can find the formal solutions near $\tau = \infty$,
then the exponential torus $\mathcal{T} $ turns out to be  isomorphic to
 $\mathbb{C}^*$, i.e., $\mathcal{T} = \{ \mathrm{diag}(c, c^{-1}), c \neq 0 \} $ and  the formal monodromy is trivial.

In fact, for the general system of that kind Balser et al. \cite{BJL} have obtained the
actual fundamental matrix solution  in terms of exponentials and Kummer's functions and as a result
they have got the Stokes matrices. The detailed calculations can be found in \cite{BJL} or \cite{MarRam}.

In our particular case (\ref{4.6}) the Stokes matrices are
\begin{equation}
\label{4.7}
St_1 =
\begin{pmatrix}
1 & 0 \\
s_1 & 1
\end{pmatrix}, \qquad
St_2 =
\begin{pmatrix}
1 & s_2 \\
0 & 1
\end{pmatrix},
\end{equation}
where
\begin{equation}
\label{4.8}
s_1 = -2 \pi i P \frac{1}{\Gamma(1+Pi) \Gamma(1-Pi)}, \qquad
s_2 = 2 \pi i P \frac{1}{\Gamma(1-Pi) \Gamma(1+Pi)}.
\end{equation}

The monodromy around the regular singular point $\tau = 0$ is $M_0 \cong \mathrm{diag} (e^{2 \pi i (iP)}, e^{-2 \pi i (iP)})$.
Since the differential Galois group $G_0$ is topologically generated by $M_0$, then $G_0 \subset G_{\infty}$ or
$G = G_{\infty}$, but this is known result for the considered systems.

It is clear from (\ref{4.7}) and (\ref{4.8}) that the identity component $G^0$ of the Galois group of the system (\ref{4.6})
is abelian if and only if $s_1 = s_2 = 0$, that is, $iP \in \mathbb{Z}$ or $\beta/\sqrt{2}\alpha \in \mathbb{Z}$.

Therefore, if
\begin{equation}
\label{4.9}
\alpha \neq  0, \quad \frac{\beta}{\sqrt{2} \alpha} \neq \mathbb{Z},
\end{equation}
the Galois group of (\ref{4.6}) is $G = G^0 = \mathrm{SL} (2, \mathbb{C})$  and the non-integrability
of the Hamiltonian system (\ref{sys}) follows from the Morales-Ramis theory.

\vspace{2ex}

{\bf Remark 1.} Alternatively, one can reduce the system (\ref{4.6}) to a particular Whittaker equation and
study its Galois group with the same end result (see \cite{M}).

\vspace{2ex}

To see that $\alpha \neq 0$ is the actual obstruction to the integrability, we proceed with the simplest
of the cases when the second condition in (\ref{4.9}) is violated, namely $\beta = 0$.

We need another particular solution.
\begin{prop}
\label{PS2}
Suppose $\alpha \neq 0, \beta = 0$. Then the system  (\ref{sys}) admits the following solution
\begin{equation}
\label{5.1}
q_2  = \frac{2 a}{\alpha}, \quad q_1 = p_1 = \frac{a}{\alpha} \exp\left(-\frac{4a^2}{\alpha} t \right),
\quad  q_3  = p_3  = \frac{i}{\sqrt{2}} q_1 , \quad  p_2  = 0.
\end{equation}

\end{prop}
The proof is immediate.

$\hfill \square$

Denote again $\xi_j = d q_j, \eta_j = d p_j, j=1, 2, 3.$ Then the variational equations (VE) along the solution
(\ref{5.1}) split nicely. Indeed, introducing the variables $v_1 = \xi_1 - \eta_1, v_3 = \xi_3 - \eta_3$ we have
\begin{eqnarray}
\label{5.2}
\dot{v}_1   & = & 2 a q_2 v_1  + 4 a q_1 \eta_2 , \nonumber \\
\dot{\eta}_2 & = & -2 a q_1 v_1 - 2 \alpha q_3^2 \eta_2 - 2 \alpha q_2 q_3 v_3 ,  \\
\dot{v}_3   & = &  \alpha q_2^2 v_3 + 4 \alpha q_2 q_3 \eta_2 . \nonumber
\end{eqnarray}
The above system  admits an integral
\begin{equation}
\label{5.3}
d \bar{H} = q_1 v_1 + q_3 v_3 := 0,
\end{equation}
which stems from the linearization of $\bar{H}$ along the solution (\ref{5.1}). With the help
of (\ref{5.3}) we remove $v_3$ from (\ref{5.2}) and obtain the system
\begin{eqnarray}
\label{5.4}
\dot{v}_1   & = & \frac{4 a^2}{\alpha} \left( v_1  + e^{-\frac{4a^2}{\alpha} t} \eta_2 \right) ,                       \\
\dot{\eta}_2 & = & \frac{2 a^2}{\alpha} e^{-\frac{4a^2}{\alpha} t}   v_1 + \frac{a^2}{\alpha} \left(e^{-\frac{4a^2}{\alpha} t}\right)^2   \eta_2.  \nonumber
\end{eqnarray}
Now, after introducing a new independent variable $z:= e^{-\frac{4a^2}{\alpha} t}$ we get the algebraic form of the above system ($' = d/dz $)
\begin{eqnarray}
\label{5.5}
v_1 '   & = & -\frac{1}{z} v_1 - \eta_2 ,                       \\
\eta_2 ' & = & -\frac{1}{2}  v_1 - \frac{z}{4} \eta_2.  \nonumber
\end{eqnarray}
Fortunately, in this case we can find the fundamental system of solutions
$$
\begin{pmatrix}
v_1 \\ \eta_2
\end{pmatrix} =
\Phi
\begin{pmatrix}
D_1 \\ D_2
\end{pmatrix},
$$
where $D_1, D_2$ are arbitrary constants and $\Phi$ is the fundamental matrix
\begin{equation}
\label{5.6}
\Phi := (\Phi^{(1)}, \Phi^{(2)}) =
\begin{pmatrix}
-\frac{z}{2} & \quad -2 \frac{e^{-\frac{z^2}{8}}}{z} - \frac{z}{2} \int \frac{e^{-\frac{z^2}{8}}}{z} d z \\[1ex]
1 & \int \frac{e^{-\frac{z^2}{8}}}{z} d z
\end{pmatrix},
\end{equation}
which in turn implies that its Galois group is solvable.

Let us see whether it is abelian.
The coefficient field of the system (\ref{5.5}) is $K := \mathbb{C} (z)$ with the usual derivation.
From the type of the solutions (\ref{5.6}) we conclude that the corresponding Picard-Vessiot extension is
$$
L := \mathbb{C} \left(z, e^{-\frac{z^2}{8}}, \int \frac{e^{-\frac{z^2}{8}}}{z} d z \right).
$$
Let $G := Gal (L/K)$ be the Galois group of (\ref{5.5}) and $\sigma \in G$, that is, $\sigma$ is a differential
automorphism of $L$ fixing $K$. Using that $\sigma \left(e^{-\frac{z^2}{8}}\right) = \delta e^{-\frac{z^2}{8}}$
and the well-known fact that $\int e^{-z^2} dz$ is not an elementary function, we have
$$
\sigma \Phi = \Phi R_{\sigma}, \quad
R_{\sigma} =
\begin{pmatrix}
1 & \gamma \\
0 & \delta
\end{pmatrix}, \, \delta, \gamma \in \mathbb{C}^* .
$$
Hence, the Galois group is represented by the matrix group
$
\Big\{
\begin{pmatrix}
1 & \gamma \\
0 & \delta
\end{pmatrix}, \, \gamma, \delta \neq 0 \Big\}
$,
which is connected, solvable, but clearly non-commutative.

\vspace{2ex}

{\bf Remark 2.} As a matter of fact, $G$ is isomorphic to the semidirect product of the additive group  and the multiplicative group
$G \cong G_a \rtimes G_m$, see e.g. Magid \cite{Magid}.

\vspace{2ex}

Therefore when $\alpha \neq 0$, the Hamiltonian system (\ref{sys}) is non-integrable by the Morales-Ramis theory.
This finishes the proof of (i).

\vspace{2ex}

{\bf Remark 3.} Note that we do not have a rigorous proof of non-integrability of the cases $\beta \neq 0$ in (\ref{4.9}).
This is because we can't find a suitable particular solution. The reasoning is as follows: since
the system (\ref{sys}) is non-integrable for the simplest and symmetric case $\beta = 0$, then it will be non-integrable for
$\beta \neq 0$ in (\ref{4.9}) again. Numerical experiments indicate chaotic behavior which suggests non-integrability.

\begin{figure}[!ht]
        \centering
        \vspace{-1ex}
%\caption{Poincar\'{e} cross sections for $\beta/(\sqrt{2}\alpha) = -1 $ and $\beta/(\sqrt{2}\alpha) = 1 $.}
        \includegraphics[width=7cm,totalheight=7cm]{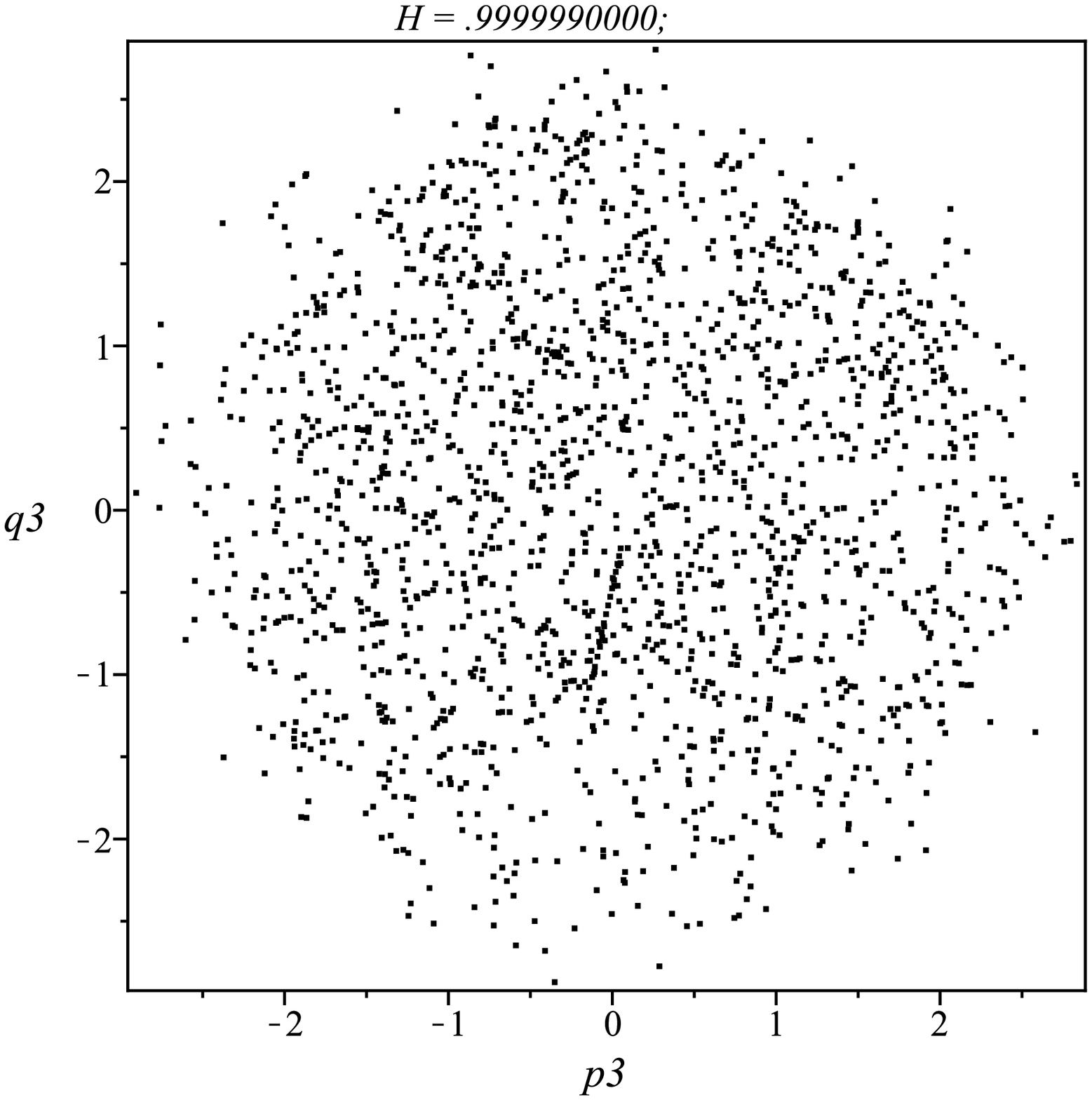}%
        \hspace{1cm}
        \includegraphics[width=7cm,totalheight=7cm]{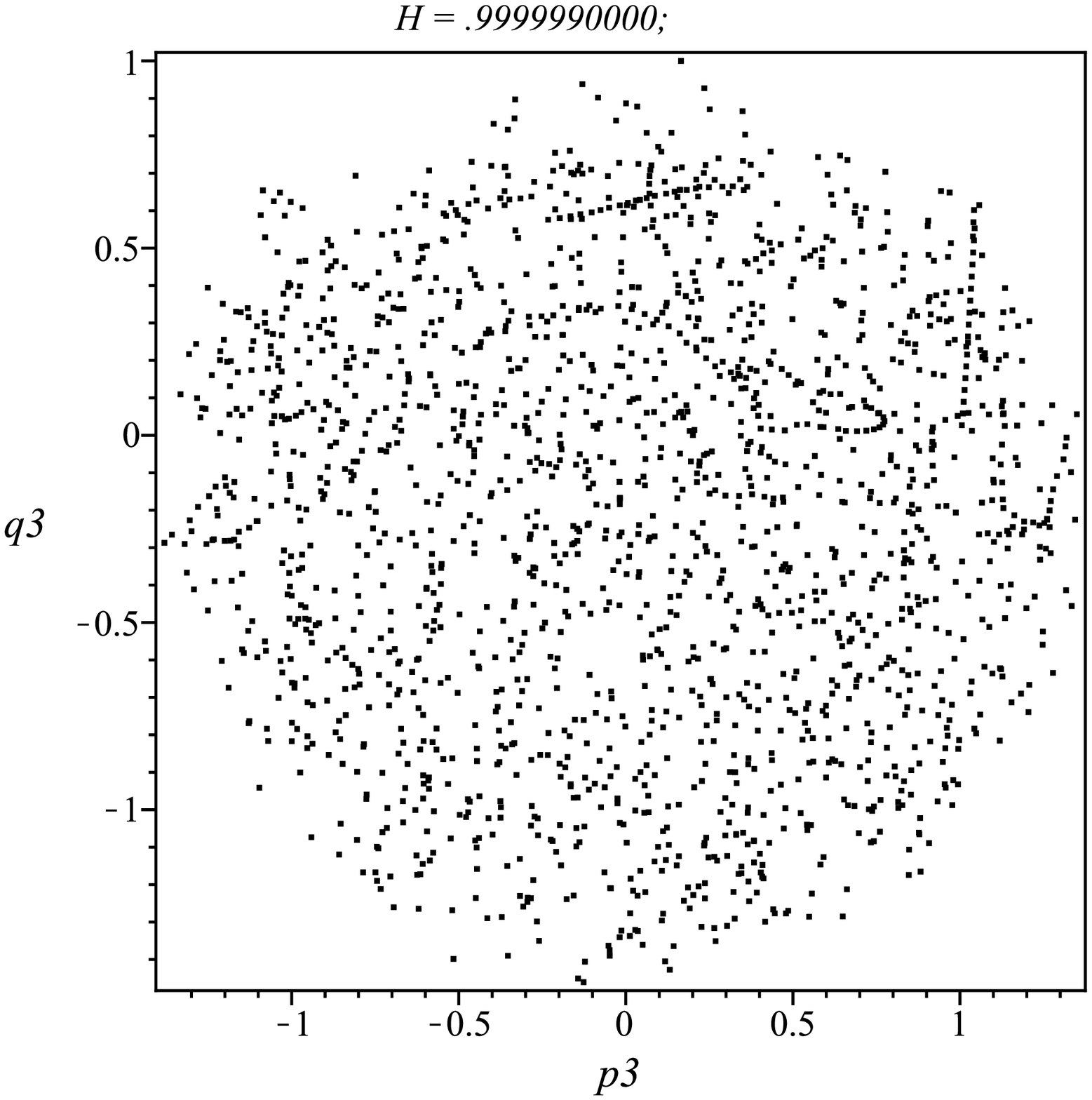}
        \vspace{5ex}
\caption{Poincar\'{e} cross sections for $\beta/(\sqrt{2}\alpha) = -1 $ and $\beta/(\sqrt{2}\alpha) = 1 $.}
     \end{figure}

\vspace{3ex}

A meticulous proof would require probably the use of higher variations \cite{MRS1,MR2}.

\vspace{3ex}

A natural question arises: what happens when $\alpha = 0, \, \beta \neq 0$?
Numerical experiments suggest a very regular behavior of the trajectories of (\ref{sys}).
Note that if any additional integral $F$ exists, it should be a real valued combination of the
generators (\ref{generators}) of the normal form since $\{H_2, F\} = 0$. After some calculations
the function $F$ (\ref{fint}) turns out to be the needed additional integral, which justifies (ii).
This proves Theorem \ref{th1}.

$\hfill \blacksquare$

\section{Concluding Remark}

In this note we study the integrability of the truncated to order four normal form of the 1 : 2 : 2 resonance.
This normal form contains too many parameters  which makes difficult the complete analysis of the problem.
Foe a generic choice of parameters we prove that the corresponding normal form is meromorphically non-integrable.
Due to the works \cite{HW,Haller} the non-integrability has a clear geometric and dynamical meaning. As in the study of other
first order resonances \cite{OC2} we use the Morales-Ramis theory.

Perhaps, the obtaining of two integrable cases is the more interesting result here. The first one (\ref{1.2}) is natural and easy,
moreover, it is KAM non-degenerated upon certain conditions on the parameters. The second one with the extra integral
(\ref{fint}) is non-trivial and can not be explained by obvious symmetry.

In fact, we could extend the non-integrability result for a larger set of parameters of the normal form, but  couldn't
succeed in finding an additional integral, so it is preferable  to state Theorem \ref{th1} in that way.
Of course, this does not mean that there are no other integrable cases.

Let us note that the algebraic approach adopted here and in \cite{OC2} can be applied to study integrability of other
resonance Hamiltonian normal forms in three degrees of freedom. cf. \cite{V2,Hansmann}.
We can remove the quadratic part of the normal form $H_2$ one way or another.
This allows us to deal with the resonances $1 : k : l$ even in the case when $k$ ( or $l$) is negative.
A systematic way how to obtain the generators of the corresponding normal forms is given in Hansmann \cite{Hansmann}.
Note that 1 : 2 : -2 and 1 : -1 : 2 are generically non semi-simple.

\vspace{3ex}

{\bf Acknowledgements.}

This work is supported by grant DN 02-5 of Bulgarian Fund "Scientific Research".

\end{document}